\documentclass[11pt]{article}
\pdfoutput=1
\usepackage{graphicx}

\usepackage{aas_macros}
\usepackage{amsmath}
\usepackage{amssymb}
\usepackage{gensymb}
\usepackage{makecell}
\usepackage{longtable}
\usepackage[font={sf}]{caption}
\usepackage{hyperref}

\usepackage[displaymath,mathlines]{lineno} 
\title{A Decade of Fast Radio Bursts}

\author{Duncan R. Lorimer$^{1,2}$}

\begin{document}

\maketitle

\begin{enumerate}
 \item Department of Physics and Astronomy, West Virginia University, P.O. Box 6315, Morgantown, WV 26506, USA
 \item Center for Gravitational Waves and Cosmology, Chestnut Ridge Research Building, Morgantown, WV 26505, USA
\end{enumerate}

\begin{abstract}
Modern astrophysics is undergoing a revolution. As detector technology
has advanced, and astronomers have been able to study the sky with
finer temporal detail, a rich diversity of sources which vary on
timescales from years down to a few nanoseconds has been found. Among
these are Fast Radio Bursts, with pulses of millisecond
duration and anomalously high dispersion compared to Galactic pulsars,
first seen a decade ago. Since then, a new research community is
actively working on a variety of experiments and developing models to
explain this new phenomenon, and devising ways to use them as
astrophysical tools. In this article, I describe how astronomers have reached
this point, review the highlights from the first decade of research
in this field, give some current breaking news, and look ahead to what 
might be expected in the next few years.
\end{abstract}

Astronomers have long predicted the existence of short-duration pulses
that might be present in high-time resolution surveys of the radio sky\cite{1}. In 2007, we published the first such example --- a bright, highly
dispersed pulse lasting 15 ms and of astrophysical but unknown origin\cite{2}. A decade on, with over 60 further examples\cite{3} of what are now
known as Fast Radio Bursts (FRBs), astronomers are now probing a
population within the landscape of transient phenomena. We now think
about the Universe both in terms of sources we see from epoch to
epoch, and transients like FRBs which provide an ever-changing
landscape rich in possibilities for future study.

The story of FRBs can be traced back to the serendipitous discovery of
pulsars by Jocelyn Bell and Anthony Hewish in 1967\cite{4}. Pulsars are
characterized by the short duration (few 10s of ms) pulses which are
extremely regularly spaced. Today, with over 2600 pulsars known\cite{5},
the currently accepted picture of a pulsar is that it is a rapidly
rotating highly magnetized neutron star which formed from the
collapsed core of a normal star after it underwent a supernova
explosion. The pulses detected on Earth are caused by the rotation of
the magnetic field which sweeps across our line of sight in much the
same way as a lighthouse. The first pulsars were found through their
individual pulses, but subsequent searches relied on the improved
sensitivity that comes from Fourier analyses of long data streams\cite{6}.

Among the first pulsars to be discovered was PSR B0531+21 in the Crab
nebula\cite{7} which was originally seen through the emission of
exceptionally bright individual pulses. The Crab and other energetic
pulsars are now known to emit what are now known as giant pulses with
a rate that follows a power-law distribution of amplitudes\cite{8}. Crab-like giant pulses could only be seen in the nearest galaxies
using the largest radio telescopes\cite{9}.

The promise of detecting pulses from energetic sources inspired a
number of searches over the years. Following Hawking's prediction in
1974\cite{10} that black holes formed early in the Universe's history
would actually be evaporating today, Rees suggested them as a
potential population of sources emitting bright radio pulses\cite{11}. Colgate also predicted radio pulses produced during supernova
explosions\cite{12}. Motivated by these predictions, Phinney and Taylor
carried out a search for bursts in data taken with the Arecibo radio
telescope\cite{13}. Although unsuccessful, this search placed the first
constraints on the rates of bursting radio sources in the
Universe. Other searches have been carried out over the intervening
years but, despite unconfirmed claims of pulses from the giant
elliptical galaxy M87\cite{14}, no further events were found\cite{14a}. 

By the end
of the 1990s, as telescope instrumentation and the availability of
high-speed computing resources improved, the interest in searching for
radio pulses was renewed\cite{15,16}.
McLaughlin et al.\cite{17} applied the single-pulse search techniques to a
large-scale survey with the Parkes radio telescope and found a new
population of sporadically emitting sources known as “Rotating Radio
Transients” (RRATs). While RRATs were quickly identified with Galactic
neutron stars, during further searches of data collected in a survey
of the Magellanic clouds, we serendipitously discovered a bright and
highly anomalous new type of pulse\cite{2}.

\begin{figure}
\includegraphics[width=\textwidth]{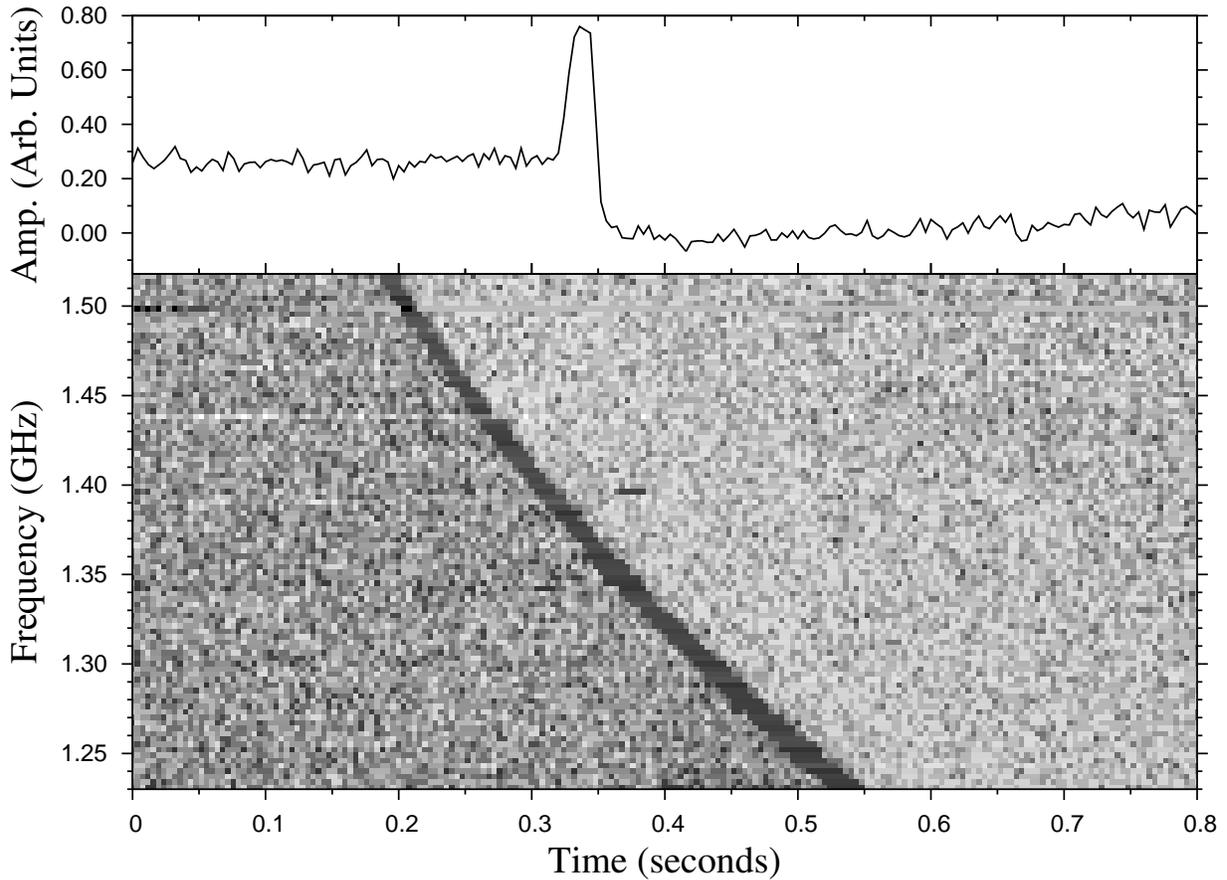}
\caption{Waterfall plot showing radio frequency versus time (lower
  panel) for the original FRB~010724 in the main beam of the
  telescope. The upper panel shows the de-dispersed pulse after
  appropriately delaying the filterbank channels to account for the
  inverse frequency squared behaviour seen below. Also evident in
  these figures is the offset level of the baseline noise prior and
  following the pulse. This is due to nature of the integrating
  circuit employed in the single-bit digitizers by this extremely
  bright pulse and was not shown in the original discovery paper
 \cite{2}. Figure credit: Evan Keane.}
\end{figure}

Fig.~1 shows the pulse we found (now known as the Lorimer burst) in
terms of radio frequency versus arrival time for a band spanning 288~MHz centred around 1374~MHz. Due to the frequency-dependent refractive
index of the ionized interstellar medium, the highest frequency
components of the signal travel faster than their lower frequency
counterparts and arrive earlier. The signature of this effect is the
inverse quadratic delay of the pulse with observing frequency\cite{4}. The
total delay across an observing band is proportional to the
dispersion measure
\begin{equation}
{\rm DM} = \int_0^d n_e \, {\rm d}l.
\end{equation} 
Here $d$ is the distance to the source and
$n_e$ is the number of electrons per unit volume along the line of
sight. It is also noticeable that the pulse gets broader towards the
lower frequencies. This is also a hallmark that is expected for a
pulse propagating through a turbulent medium and is well-known from
studies of pulsars. Using models for the electron density\cite{18},
astronomers routinely estimate the distance to pulsars and RRATs from
their observed DM values. Applying the same technique to the Lorimer
burst, it is immediately apparent that the delay is too large (by
about an order of magnitude) to be attributed to free electrons in the
Milky Way. In addition, since the location of the pulse is about three
degrees south of the Small Magellanic Cloud, we concluded that the
source was located well beyond the Milky Way and Magellanic
Clouds. Despite almost 100 hours of follow-up observations, no further
pulses were found. Making parallels with gamma-ray bursts we proposed
that this event was a prototype of a new class of astrophysical
transients, and that the rate of similar bright bursts over the sky
could be at least as large as a few hundred per day.

In spite of a great deal of effort that went into trying to identify
further examples in other archival data sets, the small fields of view
of most radio telescopes and the short duration of the bursts meant
that subsequent detections were not immediately forthcoming. Bursts
were detected that were clearly of terrestrial origin (referred to as
Perytons\cite{19}), that have subsequently been traced to emission from
microwave ovens\cite{20}. In 2012, however, a second convincing burst of
astrophysical origin from a different sky location was found in Parkes
data taken of a survey of the Milky Way\cite{21}. Like the Lorimer burst,
its DM was larger than that expected from electrons in the Milky Way
and was likely much further away. For many researchers, a
game-changing discovery in the field came in 2013 with the
announcement of four more bursts by Thornton et al.\cite{22} which clearly
revealed a population of highly dispersed anomalous pulses. At this
point the convention of naming the phenomenon Fast Radio Bursts (FRBs)
began. The designation for each FRB followed that of the gamma-ray
bursts as year, month and day of its arrival at the telescope. For
example, the FRB discovered on February 20, 2011 is called FRB
110220. Presently over 60 FRBs are known, and an up-to-date list of
source parameters can be found online\cite{3}. In addition to further
discoveries at Parkes, FRBs have also been found at Arecibo\cite{23},
Green Bank\cite{24}, Molonglo\cite{25,26}, ASKAP -- the Australian Square
Kilometre Array Pathfinder\cite{27,28} and, most recently the Canadian HI
Intensity Mapping Experiment (CHIME)\cite{29}. As shown in Fig.~2, the
anomalously high dispersion of this sample of FRBs is a defining
trait, clearly representative of a different population compared to
Galactic pulsars and RRATs.

\begin{figure}
\includegraphics[width=\textwidth]{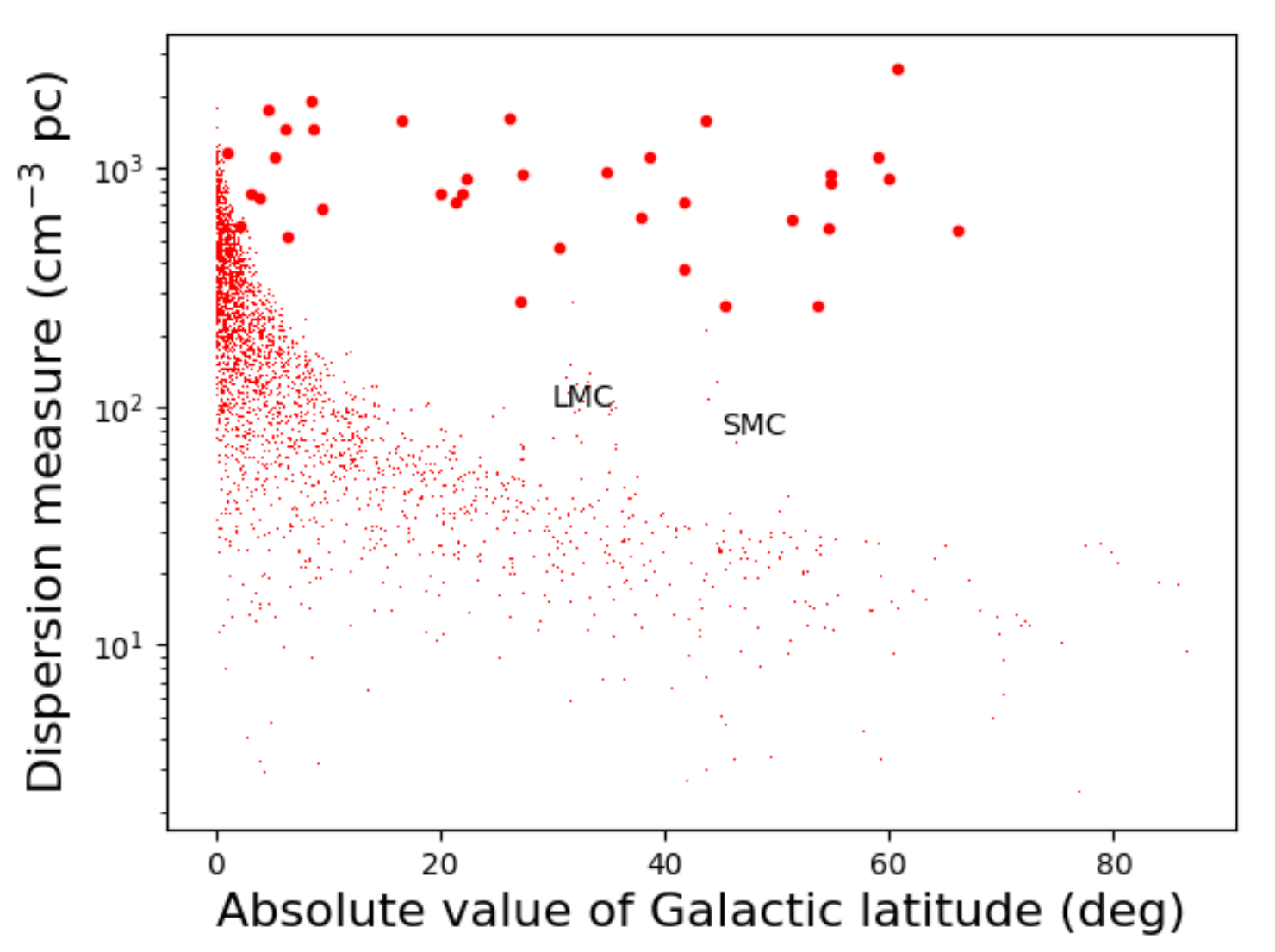}
\caption{Dispersion measure (DM) versus Galactic latitude ($b$) showing
pulsars (small dots) and Fast Radio Bursts (larger blobs). The clear
$1/\sin(b)$ dependence of DM with $b$ for pulsars as a result of the finite
size of the electron layer is evident. Also seen are faint excesses of
pulsars in the Large and Small Magellanic clouds. For FRBs, whose DMs
are not dominated by Galactic electrons, no such trend is seen.
}
\end{figure}

In an analogous way to how distances to pulsars are estimated from
their DMs, which make use of a model for the free electrons in the
Galaxy\cite{18}, redshifts of FRBs, z, can be estimated from a model of
the distribution of electrons which turns out to be dominated by those
spanning intergalactic distances. As a very crude rule of thumb\cite{30}, $z \sim$
(DM/1000 cm$^{-3}$~pc), so for typical FRBs with DMs in the range 200--2600~cm$^{-3}$~pc, we infer redshifts in the range $0.2 < z <2.6$. We
stress that this calculation uses an estimate of the redshift rather
than a direct measurement which has up to now only been possible for
one source, FRB~121102, discussed further below. However, with these
caveats in mind, for a “canonical FRB” with ${\rm DM} = 1000$~cm$^{-3}$~pc 
with a
peak flux density of a jansky and a width of 5~ms, one can infer a
co-moving distance of 3~Gpc and equivalent energy release of $10^{32}$
joules\cite{22}. For reference, the Sun emits $10^{26}$~W, so this corresponds
to the amount of energy released by the Sun in two weeks! Although
this seems like a lot, a typical Gamma-Ray Burst releases something
like $10^{45}$~joules --- more energy than the Sun will emit in its ten
billion year projected lifetime!

By the beginning of 2016, due to the lack of repeat pulses among FRBs
known at the time, in spite of hundreds of hours of follow-up time\cite{2,22}, a suspicion among many researchers was that FRBs do not
repeat and are likely one-time events. FRB~121102, originally found at
Arecibo\cite{23}, has now been observed to repeat on multiple occasions\cite{31,32,33,34}. Individual pulses are quite variable in
nature. A major implication of this discovery, however, is that for
FRB~121102, whatever is producing these bursts, it cannot be
catastrophic in nature. Models involving one-off explosions, or
stellar mergers, for this source at least are ruled out.

Further constraints on the nature of FRB~121102 have now
been found thanks to the precise localization of the source in 2017
using the Karl Jansky Very Large Array\cite{35}, and the measurement of a
significant amount of polarization in the source by the Arecibo
telescope\cite{36}. The positional determination\cite{35} allowed the
unambiguous association of FRB~121102 with a low-metallicity
star-forming dwarf galaxy at redshift $z = 0.192$\cite{37}. This is the
first, and currently only, identification of a counterpart and the
direct measurement of distance to an FRB. The luminosity distance
corresponding to this redshift is 972~Mpc. The implied isotropic
energy release of individual bursts is variable and ranges between
$10^{30}$ and $10^{33}$~joules\cite{33}.

Further radio observations using the European Very Long Baseline
Interferometry Network and Arecibo\cite{38} provide compelling evidence
for an association of FRB~121102 with a persistent radio source within
this dwarf galaxy. This could be explained as being a signature of the
emission from either the nucleus of the galaxy, or of a young pulsar
energizing a supernova remnant therein\cite{38,39}. Subsequent polarization
measurements presented by Michilli et al.\cite{36} show that the radio
emission is almost entirely linearly polarized and the amount of
Faraday rotation of the electromagnetic signal is variable, but
extremely high: in the range 130,000--150,000 radians per square
metre. Such large rotation measures are only observed in the extremely
highly magnetized environments found close to supermassive black
holes, such as Sagittarius A* at the center of our Galaxy\cite{40}.

FRB~121102 is presently unique in that it shows multiple pulses which
are sporadic in nature. Unlike the case for pulsars and RRATs, there
is no discernable periodicity. The shortest separation between bursts
currently observed is 34~ms\cite{32}. Do all FRBs repeat? While we cannot
definitively rule out the existence of repeat bursts on timescales of
years, there is now growing evidence that burst rates of the kind only
seen in FRB~121102 are the exception, rather than the rule. At the
time of writing, most FRBs have been observed for hundreds of hours,
without any evidence for repeat pulses. A recent analysis\cite{41} based
on a sample of 21 FRBs found at Parkes, if all FRBs are like 121102,
finds the probability of non-detections in the sample is less than 0.1\%. A
novel feature of the initial survey by ASKAP\cite{27,28} is that the same
fields on the sky are repeatedly observed. Each of the 20 FRBs found
in this survey has typically hundreds of hours of follow-up time
with comparable sensitivity. An analysis of these non-detections\cite{28}
also concludes that it is unlikely that they have amplitude
distributions similar to FRB~121102. Although further constraints from
other surveys and more sensitive telescopes will be more useful, it is
becoming increasingly likely that, much like gamma-ray bursts, that
there are two distinct classes of FRBs. In what follows, we will
proceed under the assumption that this is the case, i.e. that there
are both repeating FRBs and one-time FRBs.

So far, I have talked very little about what FRBs could be. This is
largely because the nature of the sources is still unclear. A
stringent constraint on the emission mechanism can be made by noting
that, for a pulse with peak flux density $S_{\rm peak}$, and width $W$
at a distance $d$
from the Earth, observing a some frequency $f$, the equivalent
temperature of a blackbody source of the same isotropic luminosity,
i.e., the brightness temperature
\begin{equation}
  T_{\rm b} = \left(\frac{S_{\rm peak}}{2\pi k}\right)
\left(\frac{fW}{d}\right)^2,
\end{equation}
where $k$ is Boltzmann's
constant. Inserting the measured values now available for FRB~121102,
we find $T_{\rm b} \sim 10^{35}$~K which clearly
implies a coherent non-thermal radiation mechanism that all
explanations summarized below must satisfy.

Very much like in the early days of gamma-ray burst astrophysics\cite{42}
there are currently more theories for the emission mechanisms than
actual FRBs known. Those put forward thus far fall broadly into three
main categories: extraterrestrial, Galactic and extragalactic. The
first of these, that FRBs represent some sort of message from
extra-terrestrial civilizations, can be largely discounted by the
preponderance of them over the whole sky. Similar arguments were made
following the discovery of pulsars\cite{4}. An intriguing possibility is
that some fraction of FRBs could be produced “accidentally” by
extra-terrestrial intelligence such as beamed emission from light
sails\cite{43}. Any source of FRBs that originates in the second category,
i.e. as natural sources from within our Galaxy, have to find a way of
explaining the extremely high DMs observed for FRBs when compared with
the pulsar population that is now well studied. Clearly FRB~121102 is
inconsistent with a Galactic origin\cite{37}. Attempts to explain the rest
of the FRBs as coming from anomalously high dispersion produced by
Galactic flare stars\cite{44}, remains less attractive due to fact that
the high implied plasma densities would not result in an inverse
frequency squared dispersion\cite{45}.

Within the extragalactic category, theories consistent with the
observed energetics involve compact objects (i.e. black holes, neutron
stars, strange stars and white dwarfs) which are necessary,
from light-travel-time considerations, to explain
the short duration of the pulses. Among the models currently suggested
are: collapsing neutron stars forming black holes\cite{46,47}, coalescing
neutron star binaries\cite{48,49}, coalescing neutron star---black hole
binaries\cite{50}, coalescing white dwarf binaries\cite{51}, primordial black
holes falling into neutron stars\cite{52}, evaporating black holes\cite{53},
giant pulses from young neutron stars\cite{54,55}, nascent and highly
energetic magnetars\cite{56,57}, white holes\cite{58}. While these scenarios
involve compact stars, non-stellar options involving cosmic strings,
large-scale defects in the structure of spacetime thought to have been
formed in early Universe, have also been proposed\cite{59,60}. Such
theories, while in principle still viable, do need to be further
developed so that they can make concrete and testable observational
predictions.  

Regardless of the origin of FRBs, predictions for ongoing and future
experiments can be made by relatively simple modeling of them as a
cosmological population. To demonstrate how a self-consistent model
can be set up, one can mimic the $S_{\rm peak}$---DM
distribution seen in the Parkes pulsars using a very simple Monte
Carlo simulation. In this model, synthetic FRBs are uniformly
distributed in co-moving volume out to $z = 2.5$. A redshift is computed
for each one and a DM is assigned which assumes a contribution of 25
cm$^{-3}$~pc from the Milky Way, $1000 z$~cm$^{-3}$~pc from the intergalactic
medium, and 150 cm$^{-3}$~pc from the host. A radio luminosity is chosen
from a normal distribution with a mean of 180~Jy~Gpc$^2$ and standard
deviation of 100~Jy~Gpc$^2$. The flux density of each FRB is then
computed. To simulate the uncertain position of each FRB within the
Parkes telescope beam, these fluxes are additionally multiplied by a
dithering factor from a Gaussian with full-width half maximum of 11
arcminutes. The result of this simulation is shown as the background
of small points in Fig.~3. This simple procedure demonstrates that the
sample requires a distribution of luminosities, and that FRBs are
unlikely to be standard candles. More detailed analyses can be found
elsewhere\cite{61,62}. With larger sample sizes and more detailed modeling
in the future, it should be possible to better constrain the
luminosity distribution and spatial distribution of the population.

\begin{figure}
\includegraphics[width=\textwidth]{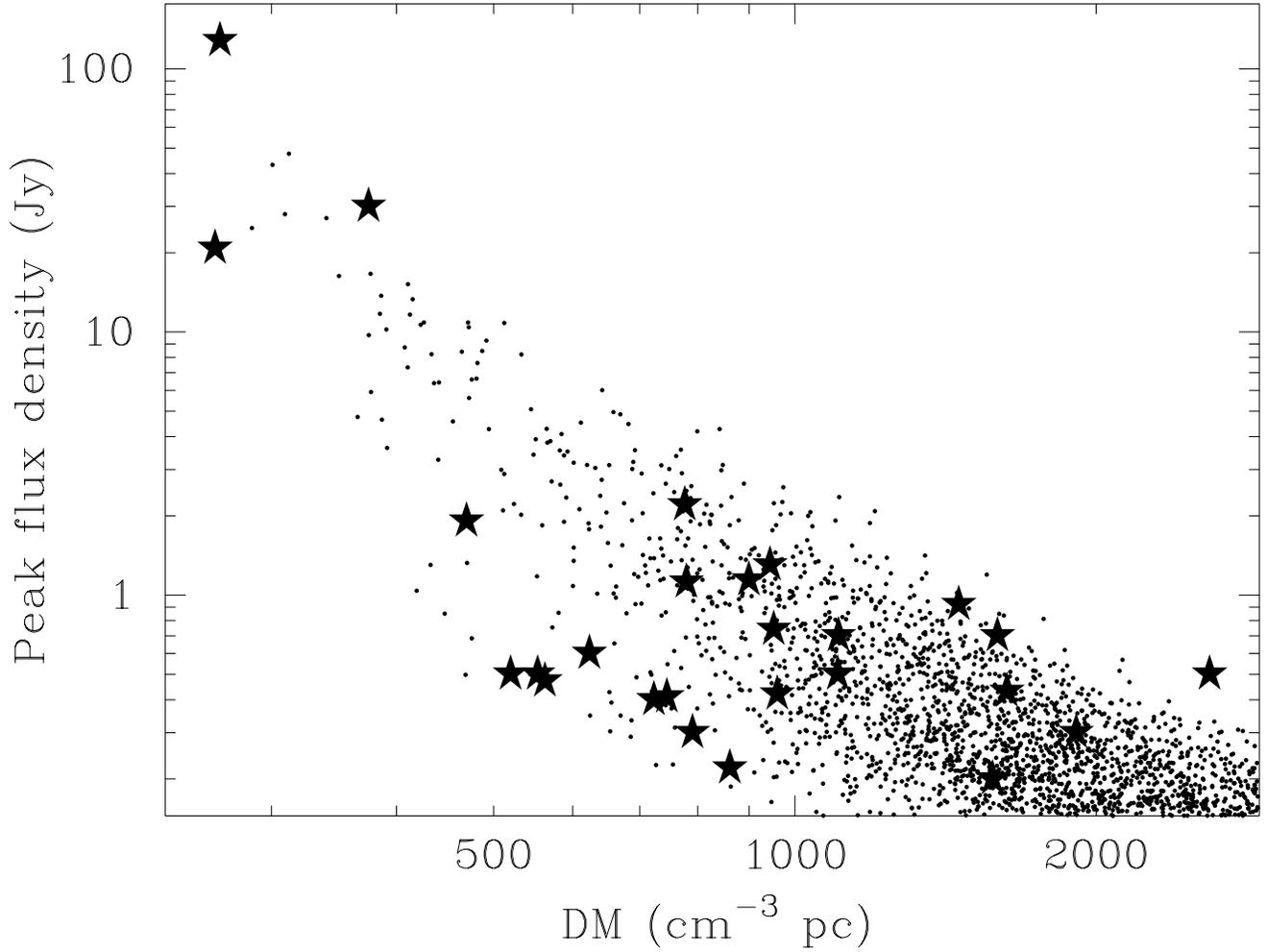}
\caption{Peak flux density versus dispersion measure (DM) for
observed (large stars) and simulated (dots) FRBs. The simulation
assumes FRBs are uniformly distributed in co-moving volume with a
Gaussian luminosity function (see text). Although the simulated events
span the DM and flux densities of the observed sample, refinements to
the model will be required to produce the correct proportion of events
as a function of DM and flux density. More sophisticated simulations
are now being carried out\cite{61,62}.
}
\end{figure}

I have summarized the discovery and salient properties of a
population of mysterious radio bursts of currently unknown origin that
appear to originate from random locations on the sky. Perhaps the most
remarkable thing about FRBs is their inferred rate of
occurrence. Based on the fields of view and time spent in surveys to
date, current estimates of their all-sky rates range between $10^3$ and
$10^4$ FRBs per sky per day\cite{22,63}. Expressed in another way, somewhere
over the sky, an FRB goes off every 30 seconds! Although all the
models we reviewed previously can be tuned to accommodate this rate,
many of them could be ruled out if we knew the distance scale for the
FRBs and, hence, their rate per unit volume in the Universe. To
emphasize the extent of the problem, FRBs are currently not too far
below the most violent known explosions in the Universe -- supernovae,
which occur somewhere in the visible universe at a rate of about one
per second. If we are missing a significant fraction of faint FRBs, or
if many of them are not beamed towards us, the rate that we infer
could be as high or even higher than the supernova rate. If, as it
seems likely, there are multiple sub-populations of FRBs
(e.g.~repeating and non-repeating sources), what are the relative
rates of these? Only through the discovery of a larger sample of these
objects will we be able to answer these questions.

Currently, new FRBs continue to be found with existing
instrumentation. ASKAP is making great inroads into the low DM
(100--1000~cm$^{-3}$~pc) population, with over 20 discoveries so far\cite{28}. ASKAP has also found the first FRB that directly probes the
line of sight of a galaxy cluster. In a 300-hr survey with ASKAP,
Agarwal et al.\cite{64} have found one bright FRB that probes the ionized
medium of the Virgo cluster. Further examples of such pulses could be
found by dedicated searches of nearby galaxy clusters\cite{65}. Very
recently, CHIME
has begun operations in the 400--800~MHz frequency band and has already
found several FRBs, including some that are visible at frequencies as
low as 400~MHz\cite{29}.

The next decade is set to be very promising for FRB hunting\cite{66}. New
instruments coming online at Arecibo, Green Bank, and elsewhere promise
to find more of these enigmatic sources. Predictions for CHIME are
extremely promising, with rates of up to dozens of sources per day
being possible\cite{67}! A new giant telescope in China -- the 
Five-Hundred-Meter Aperture Spherical Telescope (FAST) -- also recently
completed, will have unprecedented sensitivity for faint and distant
FRBs\cite{68}. At the time of writing, FAST is about to begin surveys with
a 19-beam system, while plans are underway to upgrade Arecibo's
throughput to 40 beams. Predictions for FAST and Arecibo FRBs using
the simulations described above suggest that they could see sources
out to redshift of 5 with DMs approaching $10,000$~cm$^{-3}$~pc.

A new radio telescope in South Africa, MeerKAT will soon begin
scanning the skies for FRBs\cite{meertrap} 
and an optical partner, known as MeerLICHT\cite{meerlicht}
will simultaneously observe them at optical wavelengths. Such an
experiment would probe the optical transient sky in an entirely new
way, and place further constraints on the FRB emission mechanism\cite{69}. MeerKAT is a precursor to the next generation radio telescope
known as the Square Kilometer Array (SKA) which, upon its completion
in the mid 2020s\cite{70}, would be able to locate and study FRBs to
unprecedented precision. There will be no doubt many surprises in
store as we enter the second decade of study of FRBs. I predict that
by 2030 FRBs will become standard tools to study the 
large-scale structure and magnetoionic content of the cosmos. Only
time will tell, of course, but if the first ten years is anything to
go by, FRBs still have plenty of surprises in store for astrophysics
in the next decade.

\bibliography{refs}

\end{document}